\documentclass[sigconf]{acmart}
\pdfoutput=1

\usepackage{pifont}

\newcommand{\SystemName}{PyroGuardian}
\newcommand{\AppName}{PyroPortal}
\newcommand{\HelmetWearable}{PyroHelm}
\newcommand{\NeckWearable}{PyroStrap}

\begin{document}

\title{\SystemName: An IoT-Enabled System for Health and Location Monitoring in High-Risk Firefighting Environments}

\author{Berkay Kaplan}
\email{berkay.kaplan@rutgers.edu}
\orcid{\{0000-0002-4365-7606\}}
\affiliation{
  \institution{Rutgers Business School – Newark and New Brunswick}
  \city{Piscataway}
  \state{New Jersey}
  \postcode{08854}
  \country{USA}
}

\author{Buhe Li}
\email{bl805@scarletmail.rutgers.edu}
\affiliation{
  \institution{Rutgers Business School – Newark and New Brunswick}
  \city{Piscataway}
  \state{New Jersey}
  \postcode{08854}
  \country{USA}
}

\begin{abstract}

First responders risk their lives to reduce property damage and prevent injuries during disasters. Among first responders, firefighters work with fires in residential properties, forests, or other locations where fire occurs. We built the \SystemName{} system that uses wearable modules to transmit unit information over Long Range (LoRa) to an Android tablet. The tablet runs our application, \AppName, to assign each firefighter's stats, such as body temperature, heart rate, and GPS location. \AppName{} displays this information on unit dashboards, and markers on Google Maps represent the firefighter's location and the direction they are facing. These dashboards can help the incident commander (IC) make more informed decisions on mission control operations and remove specific units whose health stats, such as oximeter and pulse, passed certain thresholds. \SystemName{} completes all these tasks at an affordable cost and in an impressive maximum range between the units and IC. In addition, \SystemName{} has various application scenarios, such as law enforcement and military operations, besides firefighting. We also conducted a sample mission inside a burning building while real firefighters watched. After the demonstration, they completed a survey on system usability and \SystemName's potential to meet their requirements.

\end{abstract}
\maketitle
\section{Introduction}

First responders operate immediately after a disaster, such as earthquakes, floods, nuclear leakages, fires, and explosions \cite{girma2020iot}. Their goal is quickly reaching the disaster point to save lives and reduce property damage \cite{girma2020iot}. Significant incidents, such as the terrorist attacks of September 11, 2001, the anthrax attacks of 2001, and the response and recovery efforts of the 2004 Southeast Asia tsunami, have emphasized the role of first responders \cite{benedek2007first}. However, they are not always safe and can suffer severe duty-related consequences.

Among first responders, firefighters frequently suffer injuries, as the NFPA 2015 National Fire Experience Survey from fire departments indicated that 68,085 firefighter injuries occurred in the line of duty in 2015 in the US \cite{haynes2015us}. Another study found that in 2019, 48 firefighters died while on duty in the US \cite{fahy2020firefighter}. The US Fire Department stated there is a fire in a residential area every 85 seconds \cite{shokouhi2019preventive}. These facts make firefighters' efficiency and safety a crucial matter for the public as firefighting is one of the most life-threatening, emotionally traumatic, and stressful occupations \cite{meina2020heart}. 

There are several factors in fire scenes, such as smoke and noise. A firefighter is deafened when they walk inside a building. They cannot see well, and radio communications are challenging \cite{microsoft_stories_2021}. Thus, an IC will not know their units' well-being. This lack of communication makes the job more difficult and dangerous. A system is needed to preserve communication between the IC and units. ICs should monitor their critical information and location to interfere in time before any injury. Relying on technological equipment can solve this problem \cite{yizhe2021design, powerdms}.

Existing solutions tackle this field with techniques like monitoring units' vitals or tracking their outdoor location. Some issues with existing solutions include the lack of GPS, preventing outdoor localization for wildfires, or vital health information monitoring. We have not found a solution that combines all these tasks into a unified dashboard for the IC. In addition, we have not found a solution that aims to keep the IC behind the safety line while allowing real-time data streaming.

To fill this gap and develop a solution while considering pricing and user convenience, we created \SystemName{}: an IoT-powered framework for firefighter mission control. \SystemName{} will warn the IC when firefighters' vitals or environment variables are dangerous. It can also stream real-time unit location and health data to the IC with a range of 610 meters.

The \SystemName{} consists of a tablet, an Android application named \AppName, an external USB LoRa adapter for \AppName, \NeckWearable{}, and \HelmetWearable{}. The IC will use \AppName{}, connected to the LoRa adapter. Each firefighter will equip one \HelmetWearable{} and \NeckWearable{}. The \HelmetWearable's casing is attachable to the exterior of the standard firefighter gas masks and helmets. The \HelmetWearable{} broadcasts sensor data, such as GPS, temperature, and 3-D inertial information, to the \AppName{} via the wireless protocol, LoRa. We used an external LoRa adapter as tablets do not have built-in LoRa. 

\AppName{} will receive additional sensor data from the \NeckWearable{} via LoRa broadcasting. The tablet has a transceiver LoRa adapter connected to its USB-C port. Once the LoRa adapter receives the sensor data, it inputs the data into the \AppName{}, displaying each unit's location and vital health information on its map. The IC can then interpret this data to navigate his officers and remove endangered units from the scene.

Another contribution of this paper includes creating a real-world fire scenario in the Illinois Fire Service Institute and observing \SystemName's efficiency during the mission. We invited 34 firefighters to watch this mission and provide feedback on our surveys. We evaluate \SystemName's system usability factor to ensure it is convenient for tradition-focused fire departments. Furthermore, the survey also asked them if \SystemName{} meets their needs. Finally, we will analyze the feasibility of our solution in other first-responder fields. Our contributions to this field include:
\begin{itemize}
  \item A novel IoT-powered solution for firefighting mission control that is cheap and easy to use.
  \item A unified dashboard that monitors the unit location and health together
  \item A user study from fire department personnel.
  \item An analysis of extending our framework to other first responder departments, such as law enforcement.
\end{itemize}

Section two gives background regarding some wireless tools and sensors integrated into \SystemName{}. The following section describes a high-level overview of \SystemName{}'s model, workflow, and configuration. The ``Implementation'' section elaborates on \SystemName{}'s sensors and breakout boards. It will explain in detail the communication between each component. In addition, ``Evaluation'' presents our results, a comprehensive user study, and \SystemName{}'s performance based on our metrics, such as cost. Then, we describe the related papers to our work. Section seven, Discussion, interprets our results and extends \SystemName{} to a broader scope. Finally, the conclusion states the future trends and ends the paper with our final remarks.

\section{Background}

The inertial measurement unit (IMU) contains three sensors, which measure acceleration through its accelerometer, orientation through the gyroscope, and magnetic direction through a magnetometer to improve the gyroscope reading \cite{ahmad2013reviews}. Developers introduced micro-electromechanical systems (MEMS) IMU with desirable features and pricing \cite{ahmad2013reviews}. It is compact with low processing power. These features increased its popularity and allowed widescale adoption \cite{ahmad2013reviews}. Researchers used IMUs for several applications, including indoor location tracking via walking \cite{hoflinger2012indoor}. We used the IMU sensor in \HelmetWearable{} to determine the direction a unit is looking.

A wireless protocol that is crucial for connecting the firefighter wearable modules to our Android application, which the IC will use, is LoRa. It is a proprietary spread spectrum modulation technique at 915.0 MHz from Semtech \cite{bor2016lora}. It can transmit information over very long distances, more specifically 10 kilometers, is resistant to effects such as multipath and Doppler, and requires low power \cite{bor2016lora, knyazev2017comparative}. Because of its increased range, this solution was suitable for ICs to communicate with the firefighters' wearable modules, as ICs can be far from disaster sights. However, it is unsuitable for high data rates, such as video feed, as its maximum data rate is 50 kbps \cite{knyazev2017comparative}. This data rate is sufficient for \SystemName{}'s sensor data.

\section{Model Overview}

\begin{figure*}[ht]
\centering
\includegraphics[width=\textwidth]{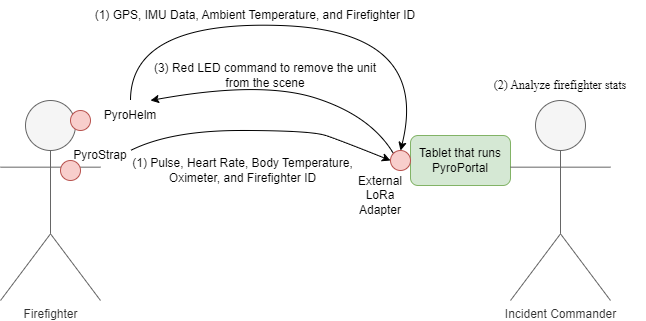}
\caption{The complete workflow of \SystemName.}
\label{figure:Shield-model}
\end{figure*}

\begin{figure}[ht]
\centering
\includegraphics[width=\columnwidth]{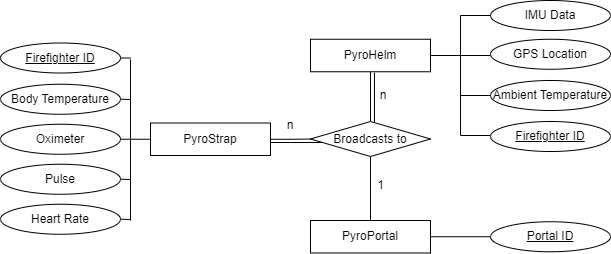}
\caption{The Chen Entity-Relationship Model of \SystemName.}
\label{figure:data-model}
\end{figure}

Our model consists of an Android application named \AppName, an arbitrary Android tablet, a USB LoRa adapter for the tablet, \NeckWearable, and \HelmetWearable. We envision the firefighter attaching the \HelmetWearable{} to his helmet as its casing can easily connect to standard gas masks and helmets. However, the user still must twist the \HelmetWearable{} to remove or insert it. We expect the firefighter to wear the \NeckWearable{} on their arm to integrate health information into our system. Figure \ref{figure:Shield-model} illustrates \SystemName{}'s workflow.

\subsection{Data Model}

\HelmetWearable{} uses sensors to get the ambient temperature, GPS, and inertial measurement unit (IMU) information. We obtain additional vital details from the \NeckWearable, such as the heart rate, blood oxygen levels, pulse, and body temperature.

Each \HelmetWearable{} and \NeckWearable{} has a hardcoded firefighter ID for the \AppName{} to recognize the firefighter's identity. Thus, ID pairing is necessary. For instance, a unit must wear the \HelmetWearable{} module with a firefighter ID 1 with the \NeckWearable{} module that has the firefighter ID 1. In addition, all wearables broadcast their firefighter ID along with sensor data with a long-range and low-power wireless protocol, LoRa, ranging 610 meters with our Adafruit boards. 

We chose LoRa as some buildings are relatively large for Bluetooth or WiFi signals. Although LoRa's data rate is minuscule, it can still fit our sensor data. \SystemName{} doesn't stream pictures or audio, as LoRa isn't the appropriate tool for it. To explain \SystemName{} data model better, figure \ref{figure:data-model} shows its Chen Entity-Relationship model.

\subsection{Data Processing at \AppName}

After the \HelmetWearable{} and \NeckWearable{} modules broadcast their messages, the LoRa transceiver adapter attached to the USB-C port of the tablet via a micro-USB to USB-C cable receives it. The adapter sends the message to the tablet using the serial output. After \AppName{} obtains the USB-related permissions from the user, it can start reading data from the USB. Afterward, \AppName{} will display the received data on the screen for the IC. The IC enters the incident address into \AppName, and the main menu becomes the Google Map view of the target building. Each firefighter will be visible on the screen. \AppName{} also warns the IC of the firefighters under injury risk when their variables pass certain thresholds, such as ambient temperatures, blood oxygen levels, or body temperature. \AppName{} uses the firefighter ID from the \HelmetWearable{} and \NeckWearable{} to identify which firefighter is broadcasting their information.

If the IC decides a unit's vitals are abnormal, they can remove them by lighting up the \HelmetWearable's red LED via the tablet's external LoRa transceiver adapter. The wearables LoRa boards are in transceiver mode, meaning they both listen to and send messages. The complete workflow of \AppName's architecture is in figure \ref{figure:Shield-model}.

Broadcasting information without authentication may not be the most secure method, as an attacker can spoof or listen to the channel. For this project's scope, we do not consider security best practices. However, we still found many LoRa authentication schemes that use techniques, such as PKI, encryption, or secure hashing, that could be integrated into \SystemName{} \cite{heeger2020analysis}.

\subsection{Analysis of Firefighter Safety with \SystemName}

Approximately one million firefighters risk their lives in the US annually \cite{smith2018assessment}. Their responsibilities and other first responders, such as law enforcement, are among the most hazardous and physically demanding \cite{smith2018assessment, marmar2006predictors}. They are often threatened and even sometimes trapped \cite{butler1998firefighter, benedek2007first}. They use radio to communicate in two ways in order to inform the IC. However, sometimes firefighters have to identify the specific location where they are present \cite{butler1998firefighter}. This identification requirement can be a dilemma if there is a wildfire in a forest. However, our GPS module covers these scenarios and provides the control team with accurate location-tracking capabilities during wildland fires.

Besides getting lost in an outdoor environment, firefighters' primary hazard is burning, which can lead to death \cite{jebakumari2021firefighter}. The burn risk can increase with units entering a building first or holding the nozzle's end \cite{jebakumari2021firefighter}. \SystemName{}'s heat-resistent ambient and body temperature sensors send data to \AppName's dashboard. When these values pass a certain threshold, their unit icons on the Google map will turn red to alert the commander. The commander can activate the \HelmetWearable{} to light its red LED up. The LED lights would signal the firefighter to return from the high-temperature zone. 

The fire service community recognizes cardiac issues as common health risks in the field \cite{fahy2020firefighter}. There are cardiovascular risks for firefighters as their duties have strenuous physical demands, sympathetic nervous system activation, increased thermal burden, and environmental exposure to smoke pollutants \cite{lefferts2021firefighter}. These factors also make cancer a well-recognized risk for retired firefighters, as the Association of Fire Fighters stated there were more than 130 firefighter cancer deaths in 2019 \cite{fahy2020firefighter}. During their operations, they experience these factors, requiring a professional to monitor their vitals. Therefore, our \NeckWearable{}'s pulse, heart rate, and oximeter sensors help the IC keep track of their units' vitals. If the commander realizes the vitals are in the danger range, they can signal the \HelmetWearable{} module to command the unit to leave the scene for rest or medical care. Although \SystemName{} does not entirely remove the injury risk, the system would mitigate it during operations and reduce cancer rates among retired units.

During indoor operations, we recognize that units will not get good GPS coverage \cite{poulose2019indoor}. However, our IMU will still operate, and its gyroscope will inform the commander of the unit's direction. While the commander will not know the unit's location inside a building, they can still partially direct units through their radio.

Another feature we are in the process of implementing using the IMU is detecting any sudden movement of the \HelmetWearable. This movement will indicate that the unit's head jerked. We can set accelerometer thresholds to catch if the movement was a jerk or if the firefighter was moving his head. This feature can detect the jerks as an indicator that an object may hit the person's head or the person fell to alert the commander. A study from Korea found that falling over or sliding during fire containment operations is very likely \cite{jo2021safety}. An object hitting the firefighter also had a high probability \cite{jo2021safety}. Thus, this IMU feature could mitigate these situations by immediately alerting the commander. This scenario can be helpful if the firefighter especially loses consciousness.

\section{Implementation}

\begin{figure*}[ht!]
    \includegraphics[width=.3\textwidth]{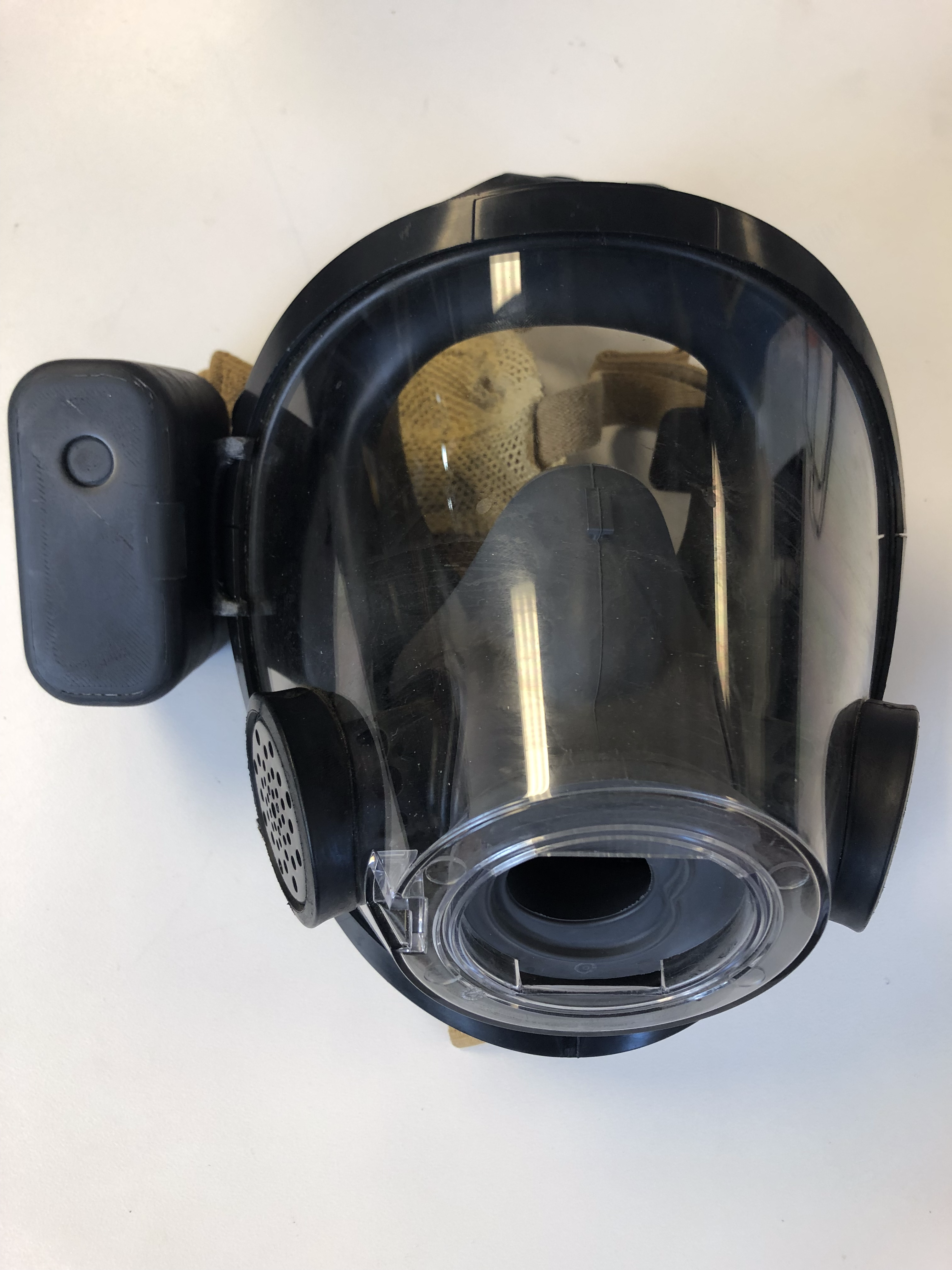}\hfill
    \includegraphics[width=.3\textwidth]{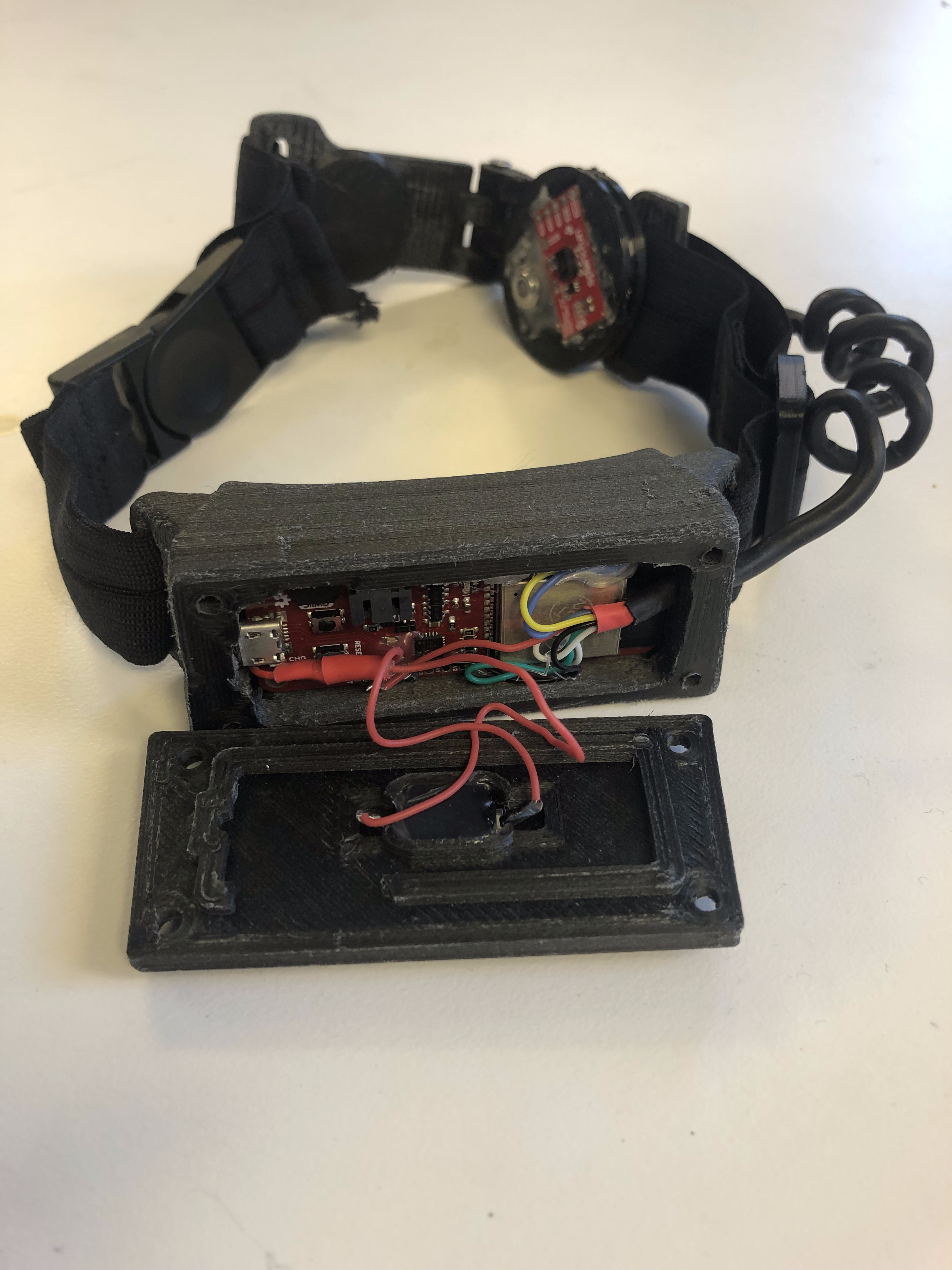}\hfill
    \includegraphics[width=.3\textwidth]{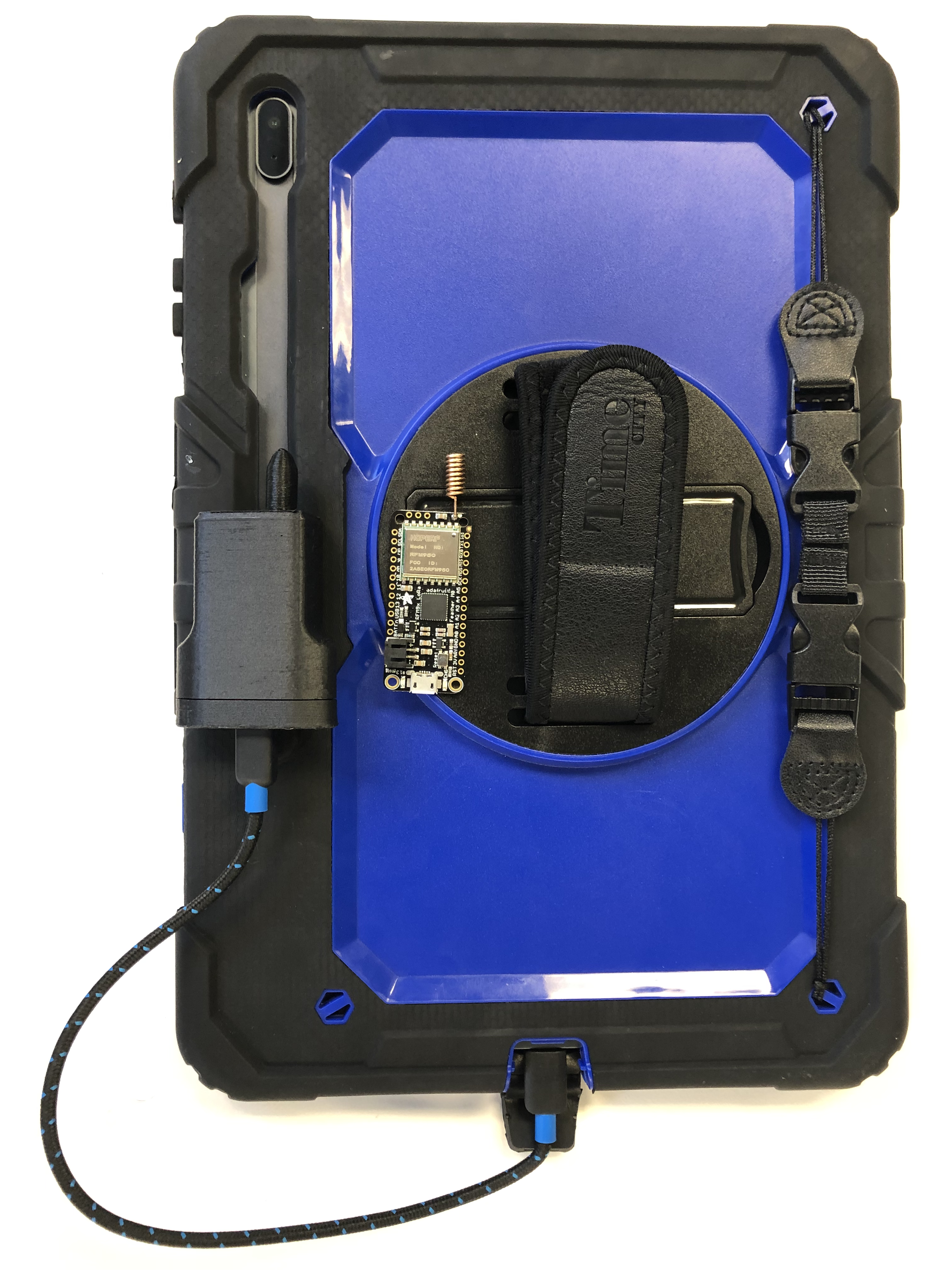}
    \caption{Elements of \SystemName. From left to right: (a) A \HelmetWearable{} module attached to a gas mask. (b) A \NeckWearable{} module with visible sensors. (c) The external LoRa adapter is on the tablet.}
    \label{figure:Shield-elements}
\end{figure*}

\begin{figure*}[ht!]
    \includegraphics[width=.48\textwidth]{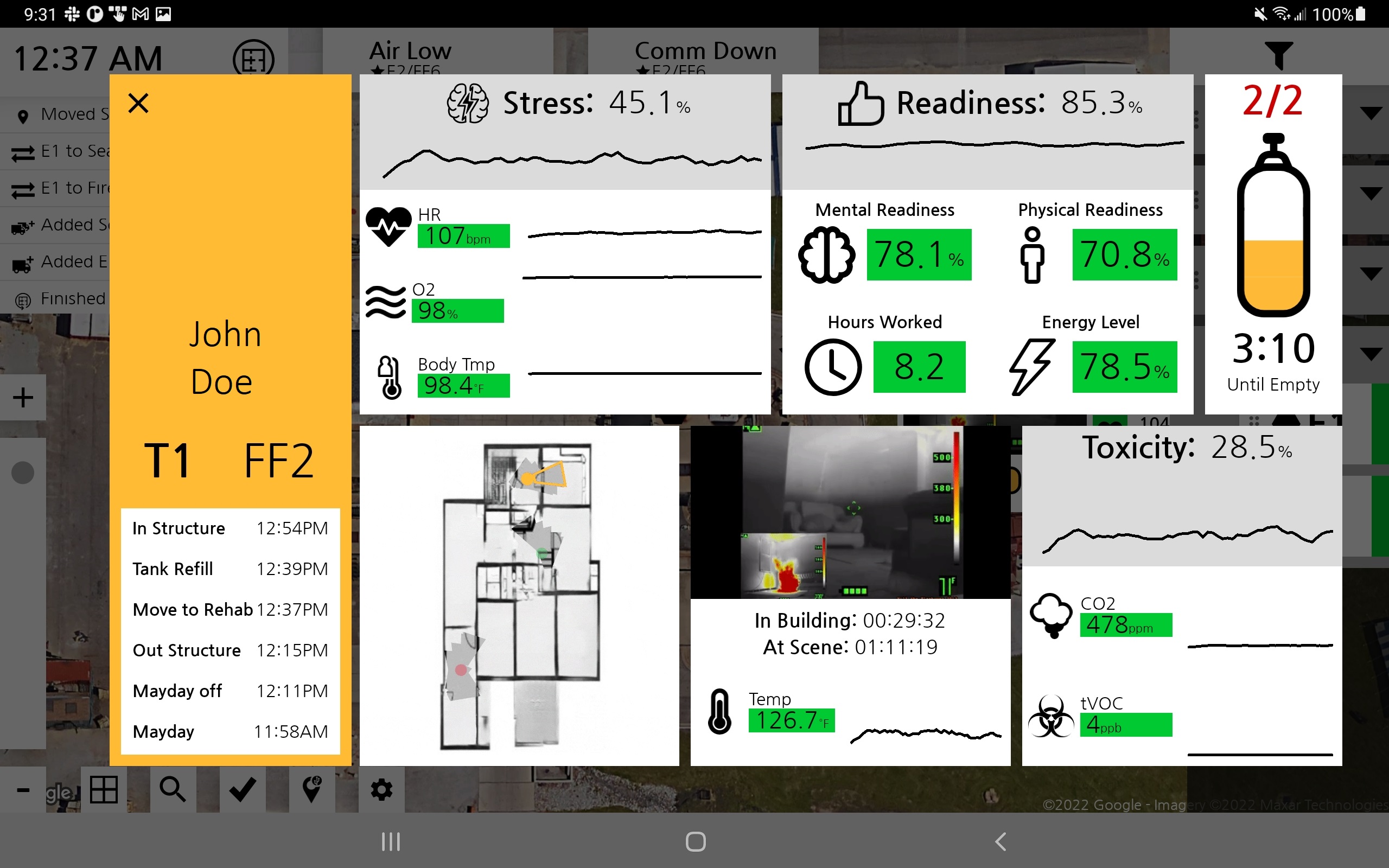}\hfill
    \includegraphics[width=.48\textwidth]{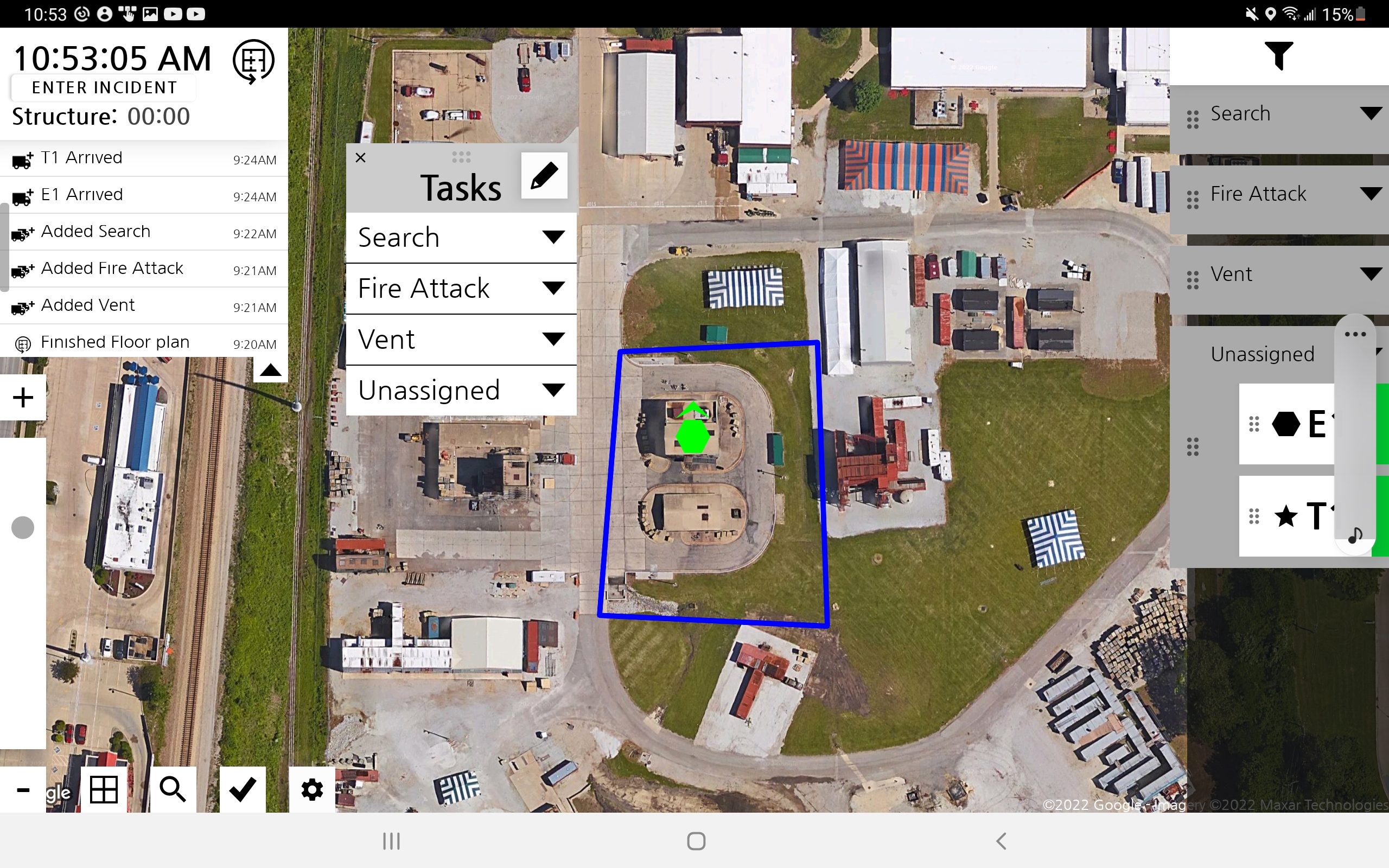}\hfill
    \caption{Screenshots of \AppName. From left to right: (a) A screenshot of a unit's dashboard in \AppName. (b) A screenshot of geofencing in \AppName.}
    \label{figure:ShieldPortal-screenshots}
\end{figure*}

We used breakout boards and sensors from vendors such as Adafruit and SparkFun. Their breakout boards use the Arduino IDE, which has an easy interface \cite{louis2016working}. This section will elaborate on the underlying hardware of \SystemName's wearable modules. It will also describe additional features in the \AppName{} applications.

\subsection{\HelmetWearable{} Sensors}

The \HelmetWearable{} has two sensors, one battery, one LoRa transceiver, and one microcontroller. The battery is a rechargeable Sparkfun model DTP603450 Polymer Li-ion, which holds 1000mAh with 3.7 volts. Along with other electrical components, it is located inside the casing to protect from the heat. 

The microcontroller is the SparkFun Thing Plus - ESP32 WROOM. It uses Qwiic I2C to connect to the sensors. Sparkfun Qwiic connectors made soldering unnecessary for connecting other sensors as it uses 4-pin JST connectors to interface other Qwiic development boards \cite{qwiic}. A view of an \HelmetWearable{} module is in figure \ref{figure:Shield-elements} section (a).

Using the Qwiic system, we daisy-chained the rest of the sensors. We added Sparkfun's SAM-M8Q GPS module to receive GPS data. Converting NMEA returns the latitude and longitude when the user is in an outdoor environment. This information helps us assign a marker to represent each firefighter's position on \AppName's Google Map screen. Next, we added Adafruit 9-DOF Absolute Orientation IMU Fusion Breakout - BNO055. It returns an Euler vector, quaternion, angular velocity vector in rad/s, acceleration vector in $m/s^{2}$ with gravity, magnetic field strength vector in micro Tesla, and a linear acceleration vector in $m/s^{2}$ without gravity \cite{townsend_2021}.

The orientation from the IMU allows us to determine which way a firefighter is facing. We update the firefighter's marker on the map with an arrow representing the direction so the commander can see where the firefighter is looking. The direction is also accurate indoors. This information would be helpful for the commander to give more informed orders through their radio. 

We have two other components that did not have a Qwiic connector. Therefore, we had to solder them to the ESP32. The LP55231 LED board still uses I2C but without a Qwiic entry. The \HelmetWearable{} module is interactive with the LP55231 board. If the IC realizes that one of the firefighter's vital information is abnormal, he can send a LoRa signal through the \AppName{} to light the LED board red. Red orders the firefighter to leave the scene at the first chance.  

We used a thermistor for temperature readings instead of a breakout board, such as the Sparkfun BME280. Our temperature sensor must endure the high heat during fires, and our current HT-NTC100K Thermistor can allow 350 Celsius maximum or -50 Celsius minimum. This sensor must also stay outside the plastic casing of \HelmetWearable{} as the \HelmetWearable's sealed container would prevent accurate temperature readings. The thermistor reaches outside through a small hole in the case and connects to the ESP32's pins. It does not use protocols such as SPI, I2C, or UART. It produces analog readings.

The ESP32 receives all the sensor data and organizes each message as a one-row CSV file with the module's firefighter ID. It broadcasts this message via the Adafruit RFM95W LoRa radio transceiver breakout board at 915 MHz. 

\subsection{\NeckWearable{} Sensors}

The \NeckWearable{} sensor is worn on the arm and has a Sparkfun Thing Plus - ESP32 WROOM controlling two sensors, one is MAX32664. It is a pulse, an oximeter, and a heart rate monitor. The second sensor is the Songhe Infrared Thermometer - MLX90614, which reads the body temperature. Besides the sensors, \NeckWearable{} uses Sparkfun model DTP603450 with 1000 mAh as its battery. As a one-row CSV message, the \NeckWearable{} broadcasts the sensor data with firefighter ID to the \AppName{} via the Adafruit RFM95W LoRa radio transceiver breakout board at 915 MHz. Unfortunately, none of the sensors uses the Qwiic connectors. Thus, some soldering was necessary. The internal view of a \NeckWearable{} module is in figure \ref{figure:Shield-elements} section (b).

\subsection{\AppName}

\AppName{} has an external adapter that is in figure \ref{figure:Shield-elements} section (c). This adapter is the transceiver Adafruit Feather M0 with an Rfm95 LoRa Radio board connected to the tablet using a micro-USB to USB-C cable. The Adafruit board would output all received messages to the serial port it is connected to. \AppName{} would constantly listen to the serial port and parse each one-row CSV message. 

Once \AppName{} reads a from the tablet's serial bus, each of the firefighters' statistical information on the screen would update. GPS information updates firefighters' position on Google Maps, while temperature, HR, and blood oxygen levels update the firefighters' stats. \AppName{} uses the unit's stats to derive specific values, such as stress for each first responder. A \AppName's screenshot in figure \ref{figure:ShieldPortal-screenshots} section (a) displays a unit's stats, which appear when an IC touches the unit's position marker. The battery level, toxicity, and readiness on the unit's dashboard are non-functional placeholders, which also include the indoor map, thermal camera, and time records in the orange box. We reserve them for future iterations. 

The IC can use the unit's stats and our derived values to make more educated decisions. If a value, such as stress, is in the danger zone, the commander can send an emergency signal to the firefighter via LoRa by pressing the emergency button. The LoRa adapter broadcasts a red LED command with the firefighter ID. The \HelmetWearable{} will constantly listen to messages and recognize the emergency command that will light the red LED to command the unit to leave the scene. We have not tested if a firefighter can see their red LED in a building on fire as visual will be limited. However, our future work will ensure that firefighters can receive a visual or another channel signal in visually limited environments.

\subsection{Geofencing}

\AppName{} allows users to draw boundaries to any area and notify the IC when a unit enters or exits this area. The user must go to the task window and click the pencil icon to enable draw mode. Once the draw mode is enabled, the user clicks on the map to choose the corners of the boundary. The corners will appear as blue dots. If a user makes a mistake, they can click the ``Undo'' button to remove their latest bounds. They can also click on ``Clear'' to remove all boundaries from the screen. When the user completes constructing their bounds, they must enter a mandatory bound name into the ``Name:'' section at the top. The user must also enter at least three corners to construct a boundary. Otherwise, \AppName{} will exit draw mode without building the bounds on the map. Finally, the user must click the same pencil icon in the tasks window to leave draw mode. The bounds will appear on the screen with blue lines, similar to figure \ref{figure:ShieldPortal-screenshots} section (b).

\subsection{Stress Calculation}

We did not find a standard to calculate stress levels, but there is work in the literature tackling this issue \cite{kim2018stress}. Several studies indicated heart rate variability (HRV) changes based on stress-induced \cite{kim2018stress}. These studies establish a link between physical pain and heart rate as pain outbreaks release adrenaline that elevates HR.

The average HR for an individual in their 20s while doing moderate-intensity physical activity is around 100 bpm, decreasing to 75 bpm until the 80s. High HRs are different for age groups. But, a high HR rule we found was subtracting one's age from 220. For instance, a 35-year-old individual should not pass 185 bpm. Since we do not know the firefighter's age, we selected a conservative 150 bpm to indicate the high-stress level in the \AppName{} dashboard by increasing the stress value.

\AppName{} also takes SpO2 levels and body temperature to calculate stress levels. A healthy O2 level range is between 95 and 100\%. If it drops below this range, individuals may experience trouble with breathing and confusion. Low blood oxygen levels may occur due to insufficient oxygen in the air or compressed breathing equipment. \AppName{} will also increase the stress level if a firefighter's oxygen level drops below 95\%.

In addition, compressed breathing air regulations, such as the National Fire Protection Association (NFPA) 1989 and CGA Grade D, also set oxygen requirements between 19.5 to 23.5 percent as 19.5\% is the lower bound for human breathing. It is essential to monitor units' oxygen levels to ensure they are not breathing in an inadequate or excessive amount.

Besides oxygen levels, \AppName{} also calculates stress with body temperature. A typical body temperature can range between 97°F (36.1°C) to 99°F (37.2°C). The standard body temperature of working firefighters can reach 39°C, and 43°C may cause death \cite{mclellan2006management}. \AppName{} will increase the stress levels if the body temperature rises from 38°C and notify the IC if it passes its critical range of 40°C.

We have yet to find metrics to update the readiness and toxicity values. Thus, they are serving as placeholders. We plan to calculate the readiness value using mental and physical readiness, hours worked, and energy levels. For future work, we will time the duration a \HelmetWearable{} was active to calculate the hours worked. However, we have not decided to calculate the other three variables under readiness. The toxicity is also essential for future work, along with CO2 and total volatile organic compounds (tVOC).

\section{Evaluation}

This section will present our user study and the performance of our solution based on the metrics we defined. This evaluation focuses on the following research questions:
\begin{itemize}
  \item RQ1: What is the cost of each element in \SystemName?
  \item RQ2: How well does \SystemName{} fulfill firefighters' needs?
  \item RQ3: How easy is it to use \SystemName?
\end{itemize}

\subsection{RQ1: \SystemName's Cost}

\begin{table*}[ht]
\centering
\begin{tabular}{| c | c | c | c |}
\hline
Breakout Board Model & Vendor & Functionality & Price \\ [0.5ex] %
\hline\hline
 Thing Plus - ESP32 WROOM & SparkFun & Microcontroller & \$22.50 \\
 RFM95W LoRa Radio Transceiver Breakout & Adafruit & LoRa & \$19.95 \\
 LED Driver Breakout - LP55231 & SparkFun & LED lights & \$10.50 \\
 SAM-M8Q - Chip Antenna & SparkFun & GPS & \$42.95 \\ 
 9-DOF Absolute Orientation IMU Fusion Breakout - BNO055 & Adafruit & IMU & \$34.95 \\ 
 3D Printer HT-NTC100K Thermistor & WINSINN & Ambient Temperature & \$2.7 \\ 
  DTP603450 Lithium Ion Battery - 1Ah & SparkFun & Power & \$10.95 \\ [1ex]
\hline
\end{tabular}
\caption{The cost of \HelmetWearable's makeup in march 2022}
\label{table:bio-extension_information}
\end{table*}

\begin{table*}[ht]
\centering
\begin{tabular}{| c | c | c | c |}
\hline
Breakout Board Model & Vendor & Functionality & Price \\ [0.5ex] %
\hline\hline
 Thing Plus - ESP32 WROOM & SparkFun & Microcontroller & \$22.50 \\
 MAX32664 & Sparkfun & Pulse, Oximeter, and Heart Rate & \$42.95 \\
 MLX90614 & Songhe & Body Temperature & \$13.88 \\ DTP603450 Lithium Ion Battery - 1Ah & SparkFun & Power & \$10.95 \\ [1ex]
\hline
\end{tabular}
\caption{The cost of \NeckWearable's makeup in march 2022}
\label{table:bio-comm_information}
\end{table*}

Firefighters must justify their purchases at a feasible price range to improve their performance metrics. Thus, it is essential to keep \SystemName{} affordable. The pricing we calculate is all items needed to run \SystemName{} on one firefighter: an Android tablet for \AppName, an External LoRa adapter, \HelmetWearable, and \NeckWearable.

We only included the prices of the electronics and not the wearables' plastic casings we printed with 3D printers. We assume a cost of \$10 for all plastic or other materials for each casing of the LoRa adapter, \HelmetWearable{}, and \NeckWearable. We listed the cost of the \HelmetWearable{} in table \ref{table:bio-extension_information}. The total cost of \HelmetWearable{} is \$154.5 along with its \$10 plastic case.

The \NeckWearable{} contains fewer sensors compared to \HelmetWearable{} and costs less. We still assume a fixed price of \$10 for non-electronics materials. We listed \NeckWearable's equipment cost in table \ref{table:bio-comm_information}. The total cost of \NeckWearable{} equipment is \$100.28.

We use an arbitrary tablet to run \AppName, but Android code is portable among other tablets. We chose Lenovo TB-J606 for \$189.99 and Adafruit Feather M0 with RFM95 LoRa Radio as the external adapter for an average use case. The board is \$34.95, and the 3.3 feet USB-C to Micro USB Cable is \$7.49, making the total cost \$242.43 with the extra \$10.

The total cost of our solution is \$497.21 with only one firefighter. Each additional firefighter will cost an extra \$254.78. 

\subsection{RQ2: Requirement Satisfaction}

\begin{table*}[ht]
\centering
\begin{tabular}{| c | c | c | c | c |}
\hline
Question & 3 & 4 & 5 & Average Score \\ [0.5ex] %
\hline\hline
 This system's capabilities will meet my requirements. & 1 & 12 & 15 & 4.5 \\  This system would be easy to use. & 4 & 12 & 12 & 4.18 \\ [1ex]
\hline
\end{tabular}
\caption{Firefighters' ratings after \SystemName's demo.}
\label{table:req_satisfaction}
\end{table*}

\begin{table*}[ht]
\centering
\begin{tabular}{| c |}
\hline
Question \\ [0.5ex] %
\hline\hline
 Regarding the question, "This system's capabilities will meet my requirements.", why did you rate it that way? \\
 Regarding the question, "This system would be easy to use.", why did you rate it that way? \\
 What functions most excite you? \\
 Which functions do you not find useful or are missing? \\ [1ex]
\hline
\end{tabular}
\caption{Survey's open-ended side questions after \SystemName's demo.}
\label{table:open_ended_questions}
\end{table*}

We invited 34 firefighters to the Illinois Fire Service Institute (IFSI) and explained to them the purpose of \SystemName. The observer firefighters came from various rural fire departments around the Champaign, Illinois, area, such as the Tuscola and Danville Fire Departments. The spectator firefighters included lieutenants, captains, fire chiefs, and deputy chiefs. IFSI lit their two-story building named Taxpayer, which contained hay. A dummy was on the second floor. The demo consisted of 2 firefighters entering Taxpayer and bringing the dummy outside to safety while equipped with \SystemName{}. 

The 34 participant firefighters watched the \AppName{} dashboard from a large-screen TV and the burning building from outside. After the two units had completed the demo, we asked the spectator firefighters to complete the UMUX-lite survey. 

The UMUX-lite survey has two questions and is on a linear scale of 1 to 5, one representing strongly disagree, three neutral, and five strongly agree. Scoring and interpreting a UMUX-lite survey is similar to a System Usability Scale (SUS) \cite{Vandereecken_2021}. Researchers normalize the average of the two survey questions' scores to get a final result between 0 and 100, where higher scores indicate better-perceived usability \cite{Vandereecken_2021}. After normalization, we obtained 86.8/100. 

We also added the four custom side questions in table \ref{table:open_ended_questions} to our UMUX-lite survey that asked the participants to explain the reasoning behind their ratings. We obtained 28 surveys in total. Finally, we interviewed each firefighter individually to learn more about their perspectives and recommendations.

The first question from our UMUX-lite survey was if \SystemName's capabilities meet their requirements. The answers we received to our survey are in table \ref{table:req_satisfaction}. We received an average score of 4.5. 

For the first and last side questions, we received recommendations to enable backup and logging for accountability and compliance for \AppName{}. \AppName{} still contains valuable information, such as the firefighter list and the incident addresses. We can integrate a cloud database and add a SIM card to the tablet to enable an internet connection. We also plan on incorporating a logging solution to keep track of each firefighter's actions in future iterations. This method can provide accurate records to ICs to improve accountability.

In the interviews, a feature that participants stated was missing was the thermal imaging streaming. Our thermal imaging is currently non-functional and a placeholder. They indicated that ICs should see the thermal video stream from firefighters' gas masks since the smoke causes the environment to be pitch-black. LoRa is not capable of video streaming. However, future technologies, such as WiFi HaLow, have higher bandwidth that may support audio or video streaming in a 1-kilometer range \cite{knyazev2017comparative}. This architecture change would reduce the theoretical range from 10 km to 1 km as HaLow has less range than LoRa \cite{knyazev2017comparative}. However, we believe 1 km would suffice.

Participants also raised privacy concerns like data security and retention practices. As indicated before, we disregard privacy topics from \SystemName{} to keep this project's scope focused on wearable modules and emergency responders.  

Finally, 6 out of 28 participants saw \SystemName{}'s cost as the most significant barrier to its adoption. Participants indicated that fire department budgets are tight, which may cause them to hesitate to adopt \SystemName{}. To reduce the cost, we thought about removing breakout boards and building our customized circuit board as we found chips cheaper than \SystemName{}'s breakout boards.

\subsection{Misc RQ2: \SystemName's Range}

We focused on the range although other metrics were available, such as \NeckWearable{} and \HelmetWearable's battery lives. We calculated the battery life to be around eight hours for each wearable.

Some disaster zones are relatively large. The maximum transmission range between the IC and the firefighter will influence the total length a unit can move before losing connection. Thus, we maximized the range between the tablet and firefighter's wearable modules with LoRa. We measured the maximum range of the LoRa Adafruit RFM95 Transceiver at 915 MHz with objects and walls between us to be 610 meters.

\subsection{RQ3: User Convenience}

The second question in our UMUX-lite survey in table \ref{table:req_satisfaction} was if \SystemName{} was easy to use. The average score we received was 4.18.  

Firefighters at the demo answered the prompt that asked which of \SystemName's features excites them the most. Our surveys' most common answers were the outdoor location-tracking features. Another firefighter indicated that this feature could drastically improve accountability with logs. Although the location feature succeeded, our participants wanted to see indoor location tracking and building floor plans.

For the second side question, our participants requested the sign-in option to preserve user-specific filters and settings. We can integrate AWS' S3 buckets to maintain settings, preferences, and logs. \SystemName's integration with the cloud would also assist with the incorporation of other systems, such as health analysis applications that can use \SystemName's vital information to provide more detailed results.

In the surveys, the majority of the participants indicated that they were impressed that one interface contains all the tasks, stats, and systems. They noted that \SystemName{} matches their response model, and having a centralized dashboard could help with incident command.

\section{Related Work}

Work suggesting wearable sensors to monitor firefighter health metrics already exists in the literature and industry \cite{meina2020heart}. However, these projects are on a much smaller scale, and none of them offer a unified dashboard that covers health, environment, and location monitoring without prior configuration. Meina et al. suggest using heart rate variability (HRV) and accelerometric data to monitor cardiovascular responses to situations \cite{meina2020heart}. Their system combines HRV and an accelerometer to measure stress levels during extreme conditions, but it does not deduce environmental information, such as a firefighter's ambient overheating. \AppName{} calculates derived values, such as stress level and toxicity.

Outdoor navigation is included in our system with GPS, as the commander can accurately track units and see the direction they're facing. This information is helpful for navigating a unit toward a particular destination. Amanatiadas et al. tackled indoor tracking and developed a system specifically for first responders \cite{amanatiadis2011intelligent}. Their system uses wearable modules, which contain a foot-mounted inertial measurement unit (IMU), a digital camera, and a radio frequency (RF) identification device for first responders \cite{amanatiadis2011intelligent}. IMU sensor utilizes the accelerometer data originating from walking patterns to perform dead-reckoning operations \cite{amanatiadis2011intelligent}. It fuses with the camera's 3D positioning algorithm and the location estimate from the Wi-Fi device \cite{amanatiadis2011intelligent}. This solution focuses on Simultaneous localization and mapping (SLAM), has a lower range with Wi-Fi, and doesn't cover health monitoring.

A work that fills the gap of vital information monitoring and still uses localization comes from O'Flynn et al. as they developed a system named SAFESENS \cite{o2018first}. It aims to provide an efficient rescue operation and first responder location tracking \cite{o2018first}. SAFESENS has eight building blocks, including UWB localization access points, an occupancy detection camera, a vital signs monitoring system, an explosive gas detector, and a firefighter tracking node \cite{o2018first}. The firefighter carries a smartphone that harvests data and sends it to a server for processing \cite{o2018first}. One of the modules is responsible for localization and uses stationary UWB anchor points \cite{o2018first}. The vital signs monitoring system performs advanced tasks, such as detecting blood composition changes \cite{o2018first}. Their system requires previously configured UWB anchor points for localization, which may not be possible in disaster scenarios. It also does not monitor environmental variables or have GPS to track firefighters outdoors. 

Instead of performing first responder localization, Fruhling et al. developed the system REaCH to solely include real-time health monitoring through first responders' wearable modules \cite{fruhling2020designing}. This system is similar to ours as they also use a dashboard to help the incident commander evaluate if their units need to leave the scene due to health complications \cite{fruhling2020designing}. However, their system does not include indoor or outdoor location tracking or geofencing.

Besides the work we found in the literature, there are also companies in this area. 3M partnered with Microsoft and developed a similar platform \cite{microsoft_stories_2021}. The platform, Florian Enroute, provides ICs a 3-D street view and vehicle tracking capabilities, which calculates their ETAs \cite{florian}. It allows ICs to use voice commands to navigate their platform and assign vehicles to specific locations \cite{florian}. Although their product has unique features, such as voice commands or geo-fencing, they do not cover vital health or environmental monitoring. Their specialization is more on location tracking, while we aim to unify these features in \SystemName. However, their voice command capability may enable ICs to perform mission control operations while driving safely and inspire us to include it in our future work. 

\section{Discussion}

This section discusses \SystemName's evaluation results, explains its further use cases, provides techniques for \SystemName's data security, and analyzes the future work items.

\subsection{Interpretation of Results}

Considering our solution with two firefighters costs approximately one thousand USD, we believe that \SystemName's cost is reasonable. However, we could not compare \SystemName's cost to any other system in the literature as we did not find their cost sections. We then looked for the pricing for similar systems in the industry. We found the Microsoft-supported 3AM Innovation's Florian system that provided features such as firefighter location, geofencing, and 3D Satellite Maps \cite{florian_3am_innovations}. We did not see Florian providing health monitoring services like \SystemName. They operate on an annual subscription basis, and only the subscription for the commander costs \$1000 \cite{florian_3am_innovations}. For each individual, it costs \$240 yearly, and each vehicle \$500 yearly \cite{florian_3am_innovations}. We believe \SystemName{} provides comparable pricing to Florian, such as our one-time fee of \$254.78 per firefighter instead of \$240 yearly, and a one-time fee of \$242.4 for only commander instead of \$1000.

Besides \SystemName's cost, we believe our 4.5 score from our UMUX-lite survey's first question is satisfactory to show that \SystemName{} covers firefighters' use case. Our 4.18 score from our UMUX-lite survey's second question is lower than the first question. We can improve the user experience in future iterations by integrating tooltips and tutorials. 

Overall, we received 86.8/100 from our UMUX-lite survey, and its benchmarks follow the SUS, which has an average score of 68 \cite{Sauro}. A SUS score between 84.1 and 100 is the best imaginable result and is in the 96-100 percentile \cite{Sauro}. The Net Promoter Score (NPS) is a metric used to measure customer satisfaction and loyalty by asking customers how likely they are to recommend a product, and our result converts to their highest score too: Promoter \cite{Sauro}. Promoters are most likely to recommend the product/website/app to a friend \cite{Sauro}.

Besides customer satisfaction metrics, we also think \SystemName's eight hours battery life and distance of 610 meters distance are acceptable for most incidents and residential buildings to allow the IC to stay safe away from the disaster zone. We can further increase this distance with a LoRa mesh network in future iterations. 

\subsection{Extended Use Cases of \SystemName}

Although \SystemName{} focused its design on firefighters, it can extend its design to several other first responder fields, such as law enforcement and the military. The \SystemName{} platform does not yet provide indoor mapping, but not every incident is inside buildings. For example, a police officer can neutralize a distressed and armed individual in venues like a street or park. The police chief can monitor the real-time location of his units behind a safety line to ensure they are outside danger zones, such as when standing too close to an armed person. 

Similarly, hot zones in military settings would require constant monitoring to ensure soldiers' safety and the mission's success. Several soldiers can get hurt during these scenarios and must receive excellent healthcare promptly. Nurses measuring hospitalized patients' health signs is a time-consuming and error-prone operation \cite{weenk2018smart}. However, injured soldiers' vitals will be on \SystemName's dashboard to help nurses gain time.

Falls in older people are a significant source of injury that can result in a disability or hospitalization \cite{pfortmueller2014reducing}. Serious injuries, such as fractures and traumatic brain injuries, occur in approximately 10\% of falls among people older than 65 \cite{pfortmueller2014reducing}. With the fall detection feature, \HelmetWearable{} can help elder care facilities. The fall detection wearable can be in the pocket of an elder. The module will not predict or stop falls or slips from happening. However, it can alert a manager if the elder is incapacitated to notify emergency personnel.

\subsection{\SystemName's Data Security}

Although we left data security out of \SystemName's scope, we still transmit sensitive health data. It is imperative to secure the transmission between the wearables and the tablet, as the LoRa wireless configuration can be vulnerable to unauthorized listeners. Another constraint is that \SystemName{} uses IoT devices that employ low-power CPUs with limited memory footprint. Thus, encryption, such as RSA, may not be suitable.

To solve this problem, in 2023, the National Institute of Standards and Technology (NIST) selected the ASCON family for lightweight cryptography standardization. \cite{lightweight_cryptography_2024}. After NIST chose ASCON algorithms for authenticated encryption with associated data and optimal hashing functionalities, organizations such as Secure-IC have already implemented those algorithms, which are compliant with the latest proposal specifications from ASCON \cite{lightweight_cryptography_2024}. We believe this can secure the data in transit on LoRa. To connect \AppName{} with the cloud, we can use conventional encryption protocols, such as TLS and HTTPS.

For data in rest, \SystemName's microcontroller, ESP32, also uses flash encryption that encrypts the contents of the ESP32's off-chip flash memory \cite{flash_encryption_for_esp32}. Once it is enabled, firmware is flashed as plaintext, and then the data is encrypted in place on the first boot. As a result, physical readout of flash will not be sufficient to recover most flash contents \cite{flash_encryption_for_esp32}. ESP32 also has a secure boot functionality that can create an even more secure environment with flash encryption \cite{flash_encryption_for_esp32}. \SystemName{} currently doesn't store user data in the tablet's file system or the cloud.

\subsection{Future Work}

GPS fixes are rare in indoor environments, so \SystemName{} does not provide indoor localization. Some techniques fuse GPS and IMU information to predict location \cite{beauregard2006pedestrian}. One such method is pedestrian dead reckoning (PDR) \cite{beauregard2006pedestrian}. It starts with a known position, and the next position displacements are added from the acceleration data \cite{beauregard2006pedestrian}. The IMU can calculate the following positions from the acceleration data, but the error will accumulate. Until the error drift becomes a significant value, the system can operate with IMU. Once the next GPS signal is received, the system can correct itself. 

Indoor location-tracking with IMU is a promising direction as researchers also managed to track objects indoors with the Kalman filter accurately \cite{poulose2019indoor}. In addition, recent advancements, such as the Vector muometric positioning system, produced centimeter-level accuracy in indoor tracking \cite{tanaka2023muometric}. We believe a standard and reliable solution for the indoor tracking problem will soon appear and integrate with \SystemName{}.

\AppName's Google Map screen can display the exterior of buildings but cannot show floor plans. If the IC knew about the floor plans, he could make more informed decisions regarding the navigation of his squads. Researchers worked on automatic indoor mapping generation techniques, which is a tedious task \cite{otero2020mobile}. Nevertheless, there are solutions for this problem, such as a handheld device that uses LiDAR or RGB-D \cite{otero2020mobile}. It is also important to note that smoke is inside the building during a fire. Popular tools, such as LiDAR, cannot function behind smoke, dust, or fog \cite{fritsche2016radar}. However, we plan to integrate ultrasonic sensors into \SystemName's architecture as they proved to be smoke-resilient during our trials.

Six hundred and ten meters would provide a reasonable range for the IC to stay behind a safety line while commanding their squads in danger zones. However, in some cases, squads may need more than this range, and we are aware that we are not utilizing the full power of LoRa, as the actual range is around 5 kilometers (km) \cite{foubert2020long}. However, other sources indicate a range of 10 km \cite{knyazev2017comparative}. We only use 23 dBm as the transmitter power, and the legal maximum is 30 dBm. The danger zone may have a large radius that spans over a km. Thus, larger hazard zones would force control teams to enter the danger zone. Using an alternative device with a higher power configuration or building a LoRa mesh network can be helpful.

The ranges of the wireless protocols were evident in our trials, but we did not focus on extending the battery lives of the two wearable modules. Some incidents, such as wildfires, may take longer than days to settle. Our current battery capacity of 1000 mAh will not cover these scenarios. Therefore, it is essential to increase the battery capacity with different methods, such as adding other rechargeable lithium-ion batteries. 

Our performance metrics only consist of two variables: cost and range. Our next iteration could focus on increasing and measuring the actual battery life with power meters. This new spec would allow \SystemName{} to give purchasers more specifications to decide whether \SystemName{} would fit their use cases. 

\section{Conclusion}

This section is the \SystemName project's closing and will present the final remarks regarding its usefulness and potential.

\subsection{Future Trends}

While the smartphone market is mature, a new set of mobile devices is trendy: wearables \cite{seneviratne2017survey}. These modules promise to increase the quality of life and fill in the gaps that smartphones could not \cite{seneviratne2017survey}. This promise will increase the variety of wearables and allow several other IoT-powered systems to emerge. We believe that sensors that power wearables will undergo a quality increase, allowing us to enhance \SystemName{} with several different features, such as indoor localization. 

Enhancements in other fields, such as wireless networks, will be beneficial to simplify the design of \SystemName{} and increase the data rate for the installation of additional modules, such as a thermal camera. A novel tool, WiFi HaLow, can achieve a data rate of up to 346.666 Mbps \cite{knyazev2017comparative}. The frequency of HaLow is 915.0 MHz in the US, with a maximum bandwidth of 16 MHz \cite{knyazev2017comparative}. Currently, we are following a particular algorithm with LoRa for broadcasting, as sometimes messages are lost due to multiple broadcasters. The increased bandwidth would allow several other modules to participate in data broadcasting. However, the downside is that LoRa has a range of 10 km while HaLow can only achieve 1 km \cite{knyazev2017comparative}. To our knowledge, HaLow prototyping boards from Adafruit or Sparkfun do not yet exist. 

Researchers around the globe are spending considerable efforts on extending battery life via several methods \cite{deyab2021improved}. Our batteries for both modules are rechargeable and have a capacity of 1000 mAh. However, fires can sometimes take longer. Furthermore, low battery life would cause the user to charge it more frequently. This frequency would hurt the user convenience and slow down adoption in the fire service field, which values tradition more than innovation \cite{powerdms}. New improvements in battery technology will make \SystemName{} require less maintenance from users. 

\subsection{Final Remarks}

\SystemName{} comprises an Android tablet and two wearable modules that sense crucial information to ensure the wearer's safety. \SystemName{} can simultaneously monitor several units' locations and provide their vital health information to the IC. One feature that makes this system unique is the usage of LoRa, as it allows the IC to stay in the safe zone while the squads can extinguish the fire. It has a maximum range of 610 meters between the IC and units.

\SystemName{} also has several other applications, such as elderly care and the military. Its wide range of sensors provides several raw variables, such as oximeter and pulse, and derives crucial values, such as stress. The wide range of sensors offers additional potential features like fall or slip detection, improved navigational monitoring via the unit's direction, and thresholds to warn the commander regarding potential injuries. We believe \SystemName{} will make a difference regarding public safety and provide the most impactful health monitoring system for first responders. 

\begin{acks}
We thank a company that preferred to remain anonymous for supporting this project with its equipment. The company's 3D casings, tablets, and microcontroller boards helped us build the physical components. The company also assisted us in conducting the user study and getting the necessary approvals to execute the demo with the Illinois Fire Service Institute. 
\end{acks}

\bibliographystyle{ACM-Reference-Format}
\bibliography{sample-base}

\end{document}